\documentclass[prb,showpacs,byrevtex,floatfix,twocolumn,superscriptaddress]{revtex4}
\usepackage{amssymb,graphicx,afterpage,amsmath,color}
\begin{document}
\title{\boldmath Four-coloured Spin-wave Excitations in Multiferroic Materials \unboldmath}

%
%
%
\author{I. K\'ezsm\'arki}
\affiliation{Department of Physics, Budapest University of
Technology and Economics, 1111 Budapest, Hungary}
\affiliation{Condensed Matter Research Group of the Hungarian
Academy of Sciences, 1111 Budapest, Hungary}
\author{D. Szaller}
\affiliation{Department of Physics, Budapest University of
Technology and Economics, 1111 Budapest, Hungary}
\author{S. Bord\'acs}
\affiliation{Quantum-Phase Electronics Center and Department of
Applied Physics, University of Tokyo, Tokyo 113-8656, Japan}
\author{H. Murakawa}
\affiliation{RIKEN Center for Emergent Matter Science (CEMS), Wako
351-0198, Japan}
\author{Y. Tokura}
\affiliation{Quantum-Phase Electronics Center and Department of
Applied Physics, University of Tokyo, Tokyo 113-8656, Japan}
\affiliation{RIKEN Center for Emergent Matter Science (CEMS), Wako
351-0198, Japan}
\author{H. Engelkamp}
\affiliation{High Field Magnet Laboratory, Institute for Molecules
and Materials, Radboud University, 6525 ED Nijmegen, The
Netherlands}
\author{T. R{\~o}{\~o}m}
\affiliation{National Institute of Chemical Physics and Biophysics,
Akadeemia tee 23, 12618 Tallinn, Estonia}
\author{U. Nagel}
\affiliation{National Institute of Chemical Physics and Biophysics,
Akadeemia tee 23, 12618 Tallinn, Estonia}
\date{\today}
\begin{abstract}
The optical magnetoelectric effect, which is an inherent attribute
of the spin excitations in multiferroics, drastically changes their
optical properties compared to conventional materials where
light-matter interaction is expressed only by the dielectric
permittivity and magnetic permeability. Our polarized absorption
experiments performed on multiferroic Ca$_2$CoSi$_2$O$_7$ and
Ba$_2$CoGe$_2$O$_7$ in the THz spectral range demonstrate that such
magnetoeletric spin excitations show quadrochroism, i.e. they have
different colours for all the four combinations of the two
propagation directions (forward or backward) and the two orthogonal
polarizations of a light beam. We found that quadrochroism can give
rise to peculiar optical properties, such as one-way transparency
and zero-reflection of these excitations, which can open a new
horizon in photonics. One-way transparency is also related to the
static magnetoelectric phenomena, hence, these optical studies can
provide guidelines for the systematic synthesis of new materials
with large dc magnetoelectric effect.
\end{abstract}
\maketitle

Nature offers a plethora of dichroic materials whose colour is
different for two specific polarizations of the transmitted light.
Dichroism generally appears in media which are not isotropic and
provides information about their symmetry. Linear dichroism
(absorption difference for two orthogonal linear polarizations of a
light beam) emerges in materials where the symmetry is lower than
cubic, while circular dichroism (absorption difference for the two
circular polarizations) is observed in materials with finite
magnetization or chiral structure. In all these cases the four
transverse wave solutions obtained from the Maxwell equations for a
given axis of light propagation group into two pairs, where each
pair contains two counter-propagating waves ($\pm$\textbf{k})
characterized by the same absorption
coefficient.\cite{Azzam1979,Barron2004}

In materials with simultaneously broken spatial inversion and time
reversal symmetry this two-fold $\pm$\textbf{k} directional
degeneracy of the Maxwell equations can be lifted and each of the
four waves is absorbed with a different
strength.\cite{Hornreich1968,O'Dell1970,Arima2008_3,Cano2009,Szaller2013}
Hereafter, we will refer to this phenomenon as quadrochroism and the
corresponding materials as quadrochroic or four-coloured media.
Following the early prediction of $\pm$\textbf{k} directional
anisotropy\cite{O'Dell1970,Baranova1979,Barron1984} and experimental
observation\cite{Rikken1997,Rikken2002} of weak directional effects,
recent optical studies on multiferroic materials report about strong
directional dichroism in the
GHz-THz\cite{Cano2009,Kezsmarki2011,Bordacs2012,Takahashi2012,Takahashi2013}
and visible spectral range\cite{Arima2008,Arima2008_2,Arima2009} as
a hallmark of quadrochroism.


\section{Results}

\textbf{Quadrochroism generated by the optical magnetoelectric
effect.} Here, we argue that the emergence of quadrochroism is not a
fortuitous issue but an inherent property of magnetoelectric
multiferroics. In this class of materials, the coupling between the
electric and magnetic states leads to the optical magnetoelectric
effect described by the dynamical response functions
$\hat{\chi}^{me}(\omega)$ and
$\hat{\chi}^{em}(\omega)$.\cite{O'Dell1970,Cano2009,Kezsmarki2011,Bordacs2012,Arima2008_3,Miyahara2011}
Consequently, an oscillating magnetization and polarization
($M_i^\omega$ and $P_i^\omega$) are induced by the electric and
magnetic components of light ($E_i^\omega$ and $H_i^\omega$),
respectively, besides the conventional terms arising from the
dielectric permittivity ($\hat{\epsilon}$) and magnetic permeability
($\hat{\mu}$);
$M_i^\omega$$=$$[\mu_{ij}(\omega)-1]H_j^\omega+\sqrt{\epsilon_0/\mu_0}\chi^{me}_{ij}(\omega)E_j^\omega$
and
$P_i^\omega$$=$$\epsilon_0[\epsilon_{ij}(\omega)-1]E_j^\omega+\sqrt{\epsilon_0\mu_0}\chi^{em}_{ij}(\omega)H_j^\omega$.
Solving the Maxwell equations with these generalized constitutive
relations yields the following form of the complex refractive index
for a given polarization ($E_i^{\omega}$,
$H_j^{\omega}$):\cite{Kezsmarki2011,Miyahara2012}
\begin{equation}
N_{\pm}(\omega)\approx\sqrt{\epsilon_{ii}(\omega)\mu_{jj}(\omega)}\pm\frac{1}{2}[\chi^{me}_{ji}(\omega)+\chi^{em}_{ij}(\omega)],
\end{equation}
where $\pm$ signs correspond to wavevectors $\pm$\textbf{k}. The
second term in this formula explicitly shows the key role of the
optical magnetoelectric effect in generating quadrochroism by
lifting the $\pm$\textbf{k} degeneracy of the Maxwell equations. The
derivation of Eq.~1 for magnetoelectric materials of various
symmetries is given in the Methods section.

The optical magnetoelectric effect is exclusively generated by such
transitions where both the electric- and magnetic-dipole moments
induced by an absorbed photon are finite:
\begin{align}
\chi^{me}_{ji}(\omega)&=
\frac{2}{V\hbar}\sqrt{\frac{\mu_0}{\epsilon_0}}\sum_n\left[
\frac{\omega_{no}\Re\{\langle 0|m_j|n\rangle\langle
n|p_i|0\rangle\}}{\omega_{no}^2-\omega^2-2i\omega\delta}\right.\nonumber\\
&\left.+\frac{i\omega\Im\{ \langle 0|m_j|n\rangle \langle
n|p_i|0\rangle\}}{\omega_{no}^2-\omega^2-2i\omega\delta}\right]\triangleq\chi^{\prime}_{ji}(\omega)+\chi^{\prime\prime}_{ji}(\omega),
\end{align}
i.e. the magnetic ($m_j$) and electric ($p_i$) dipole operators
simultaneously must have non-vanishing matrix elements between the
ground state $|0\rangle$ and the excited states $|n\rangle$
separated by $\hbar\omega_{no}$ energy. Here, $V$ is the volume of
the system, $\epsilon_0$ and $\mu_0$ are respectively the
permittivity and the permeability of the vacuum.
$\chi^{\prime}_{ji}(\omega)$ and $\chi^{\prime\prime}_{ji}(\omega)$
are the sum of terms with the real ($\Re$) and the imaginary ($\Im$)
parts of the matrix element products, respectively. The former
changes sign under time reversal, while the latter remains
invariant. Since the two cross effects are related by the Kubo
formula according to
$\chi^{me}_{ji}(\omega)=\chi^{\prime}_{ji}(\omega)+\chi^{\prime\prime}_{ji}(\omega)$
and
$\chi^{em}_{ij}(\omega)=\chi^{\prime}_{ji}(\omega)-\chi^{\prime\prime}_{ji}(\omega)$,
in Eq.~1 the time-odd $\chi^{\prime}_{ji}(\omega)$ is solely
responsible for the  $\pm$\textbf{k} directional anisotropy;
$N_+(\omega)-N_-(\omega)=\chi^{me}_{ji}(\omega)+\chi^{em}_{ij}(\omega)=2\chi^{\prime}_{ji}(\omega)$.

When the optical magnetoelectric effect is sufficiently strong, a
four-coloured material can become fully transparent for a given
propagation direction, while it still absorbs light beams traveling
in the opposite direction. We define the magnetoelectric ratio for a
given transition as $\gamma$$\triangleq$$\frac{\langle
n|m_j|0\rangle}{\langle n|p_i|0\rangle}$. For a magnetoelectric
resonance separated from other excitations the one-way transparency,
that is
$\alpha_-(\omega)$$\equiv$$\frac{2\omega}{c}\Im\{N_-(\omega)\}$$=$$0$
while
$\alpha_+(\omega)$$\equiv$$\frac{2\omega}{c}\Im\{N_+(\omega)\}$$\neq$$0$
or vice versa, can occur for a given polarization $E_i^{\omega}$ if
\begin{equation}
\gamma=\pm\frac{c}{\sqrt{\epsilon^{\infty}_i}},
\end{equation}
where $c$ is the speed of light in vacuum, $\epsilon^{\infty}_i$ is
the dielectric permittivity due to optical transitions at higher
frequencies. Note that one-way transparency is only possible when
$\gamma$ is purely real, i.e. the interference between the magnetic-
and electric-dipole matrix elements has to be perfectly constructive
or destructive. Correspondingly, the $\pm$ signs refer to the two
cases when the polarization and the magnetization induced by the
electromagnetic field oscillate with zero and $\pi$ phase
difference, respectively. The condition in Eq.~3 can be equivalently
expressed by the susceptibilities in the frequency region of the
individual magnetoelectric transition, namely
$\chi^{\prime\prime}_{ji}(\omega)$ has to vanish and the ratio of
the dynamical magnetic and electric susceptibility needs to be equal
to $1/\epsilon^{\infty}_i$. Note that in CGS units Eq.~3 has the
form $\gamma$$=$$\pm\frac{1}{\sqrt{\epsilon^{\infty}_i}}$, hence,
the magnetic- and electric-dipole matrix elements must be of the
same order of magnitude to approach the one-way transparency. (For
details see the Methods section.)

\begin{figure}[t!]
\includegraphics[width=3.4in]{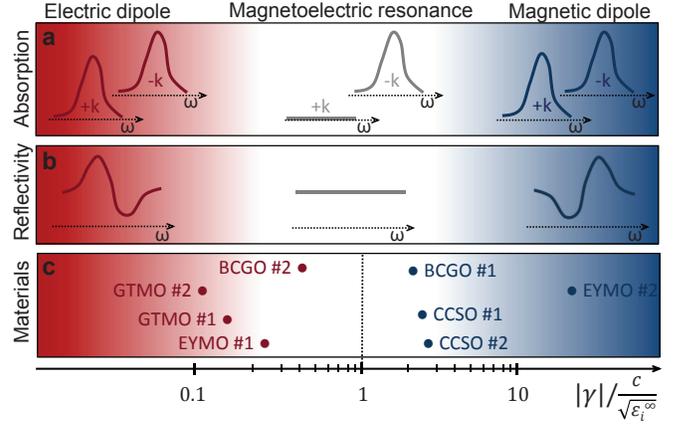}
\caption{\textbf{$\mid$ One-way transparency at magnetoelectric
resonances.} Optical properties characteristic to pure
electric-dipole, mixed magnetoelectric and pure magnetic-dipole
excitations corresponding to different values of the magnetoelectric
ratio, $\gamma$. \textbf{a,} Counter-propagating beams are absorbed
with the same strength in case of pure electric (red region) and
magnetic (blue region) dipole transitions as illustrated by the two
absorption peaks labeled with $+$\textbf{k} and $-$\textbf{k} (for
clarity the two peaks are shifted relative to each other). When
approaching the limit of ideal magnetoelectric resonance (white
region), where $\left|\gamma\right|=c/\sqrt{\epsilon^{\infty}_i}$,
the material tends to show one-way transparency. \textbf{b,}
Electric- and magnetic-dipole transitions emerge with opposite line
shapes in the reflectivity spectrum, while an ideal magnetoelectric
resonance remains hidden. \textbf{c,} Magnetoelectric ratio for
magnon modes with strong $\pm$\textbf{k} directional anisotropy
observed in multiferroic Ca$_2$CoSi$_2$O$_7$ (CCSO),
Ba$_2$CoGe$_2$O$_7$ (BCGO), Gd$_{0.5}$Tb$_{0.5}$MnO$_3$
(GTMO)\cite{Takahashi2013} and Eu$_{0.55}$Y$_{0.45}$MnO$_3$
(EYMO).\cite{Takahashi2012} For the mode assignments see the main
text and Fig.~4. The $\gamma$ values and their field dependence for
the different modes are given in the Supplementary Fig.~S1.}
\end{figure}

Although directional anisotropy is not manifested in the
reflectivity of a vacuum-material interface at normal incidence, the
one-way transparency of a bulk material can still be accompanied by
peculiar behaviour of the normal-incidence reflectivity, namely the
magnetoelectric resonance remains hidden in the reflectivity
spectrum, $R(\omega)$. Purely electric- and magnetic-dipole
transitions appear in the reflectivity spectrum with opposite line
shapes according to the Fresnel formula
$R$$=$$\left|\frac{1-\sqrt{\frac{\mu_{jj}}{\epsilon_{ii}}}}{1+\sqrt{\frac{\mu_{jj}}{\epsilon_{ii}}}}\right|^2$
as schematically shown in Fig.~1. Thus, the relative strength of
magnetic and electric susceptibilities for a given resonance can be
determined from the shape of $R(\omega)$ near the resonance. When
the condition $\left|\gamma\right|$$=$$c/\sqrt{\epsilon^{\infty}_i}$
is fulfilled the reflectivity spectrum is featureless at the
resonance with the constant value of
$R$$=$$\left|\frac{\sqrt{\epsilon^{\infty}_i}-1}{\sqrt{\epsilon^{\infty}_i}+1}\right|^2$
as if there was no resonance present in that frequency range. (See
the Methods section.)
\begin{figure}[t!]
\includegraphics[width=2.8in]{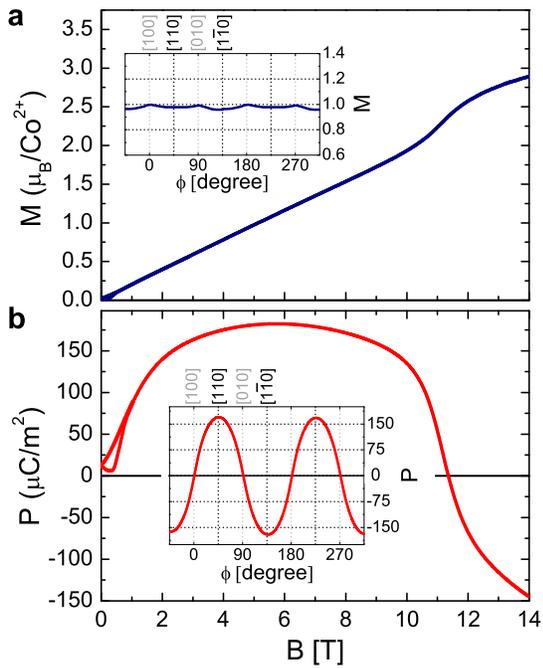}
\caption{\textbf{$\mid$ Magnetoelectric properties of multiferroic
Ca$_2$CoSi$_2$O$_7$ at $T$$=$$2$\,K.} Field dependence of
\textbf{a,} the magnetization (M) and \textbf{b,} the ferroelectric
polarization induced along the tetragonal [001] axis (P) for
magnetic fields parallel to the [110] direction. There is a
hysteresis in both quantities for fields $<$1\,T. The insets display
the angular dependence of the magnetization and polarization when a
magnetic field of B$=$5\,T is rotated within the tetragonal plane.}
\end{figure}

\begin{figure*}[t!]
\includegraphics[width=6.6in]{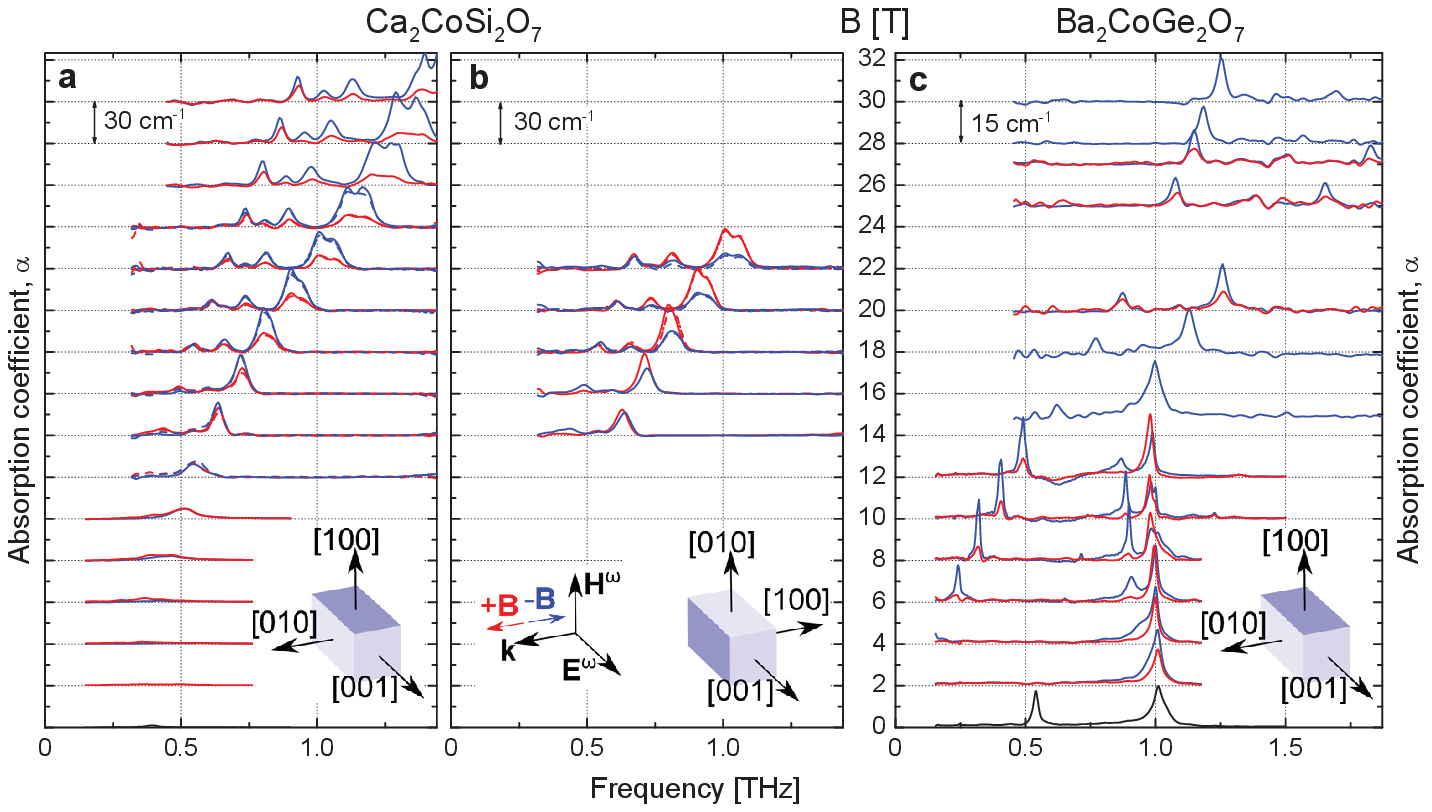}
\caption{\textbf{$\mid$ Directional dichroism of the magnon modes in
Ca$_2$CoSi$_2$O$_7$ and Ba$_2$CoGe$_2$O$_7$.} Absorption spectra
measured in Faraday configuration, i.e. for
\textbf{B}$\parallel$\textbf{k}, in magnetic fields B$=$0-30\,T at
T$=$2\,K. The spectra are shifted vertically proportional to the
magnitude of the field and the distance between horizontal grid
lines is indicated in each panel. The axis of the static magnetic
field together with the polarization and the propagation direction
of light, common for all spectra in the figure, are shown in panel
\textbf{b}, while the corresponding sample orientation is specified
in each panel. In panel \textbf{a} and \textbf{c,} the polarization
configuration is {\bf E$^\omega$}$\parallel$[001], {\bf
H$^\omega$}$\parallel$[100] and {\bf k}$\parallel$[010], while in
panel \textbf{b,} the sample is rotated by $\pi/2$ around the [001]
axis. Red and blue spectra correspond to light propagation along and
opposite to the magnetic field, respectively. In both materials,
some of the modes exhibit strong directional ($\pm$\textbf{k})
dichroism, absorption difference for counter-propagating light
beams, almost realizing one-way transparency. Measurements on
Ca$_2$CoSi$_2$O$_7$, shown in panel \textbf{a} and \textbf{b}, have
been repeated after $\pi$ rotations of the sample around the [100],
[010] and [001] axes. The corresponding spectra plotted with dashed
lines nearly coincide with the original ones in the whole frequency
range.}
\end{figure*}

\textbf{Multiferroic character of Ca$_2$CoSi$_2$O$_7$.} This
compound crystallizes in a non-centrosymmetric tetragonal
$P\overline{4}2_1m$ structure\cite{Hagiya1993,Jia2006} where
Co$^{2+}$ cations with S=3/2 spin form square-lattice layers stacked
along the tetragonal [001] axis. The static magnetic and
magnetoelectric properties of
Ca$_2$CoSi$_2$O$_7$,\cite{Akaki2009,Akaki2009_2} as shown in Fig.~2,
resemble to the properties of other multiferroic compounds from the
same family such as
Ba$_2$CoGe$_2$O$_7$\cite{Murakawa2010,Yi2008,Romhanyi2011,Yamauchi2011,Perez-Mato2011}
and Sr$_2$CoSi$_2$O$_7$.\cite{Akaki2012} This material shows an
antiferromagnetic ordering below $T_N$$=$$5.7$\,K with a small
ferromagnetic component of the spins (M) lying in the tetragonal
plane.\cite{Zheludev2003,Hutanu2012} Below $T_N$, magnetic fields
applied along the [110] (or [1$\overline{1}$0]) axis of the the
tetragonal plane induce ferroelectric polarization along the [001]
direction (P), which changes sign at $B$$\approx$$11$\,T accompanied
by an anomaly in the magnetization. The rotation of $B$$=$$5$\,T
magnetic field within the tetragonal plane results in a nearly
sinusoidal modulation of P with zero crossing for fields pointing
along the [100] and [010] axes. The variation of M is less than
$5\%$ implying that in-plane magnetic anisotropies, due to e.g.
small orthorhombicity of the crystal structure,  become negligible
in this field range. On this basis we expect that the magnetic
symmetry of the material can be approximated by the
$mm^\prime2^\prime$ ($m^\prime m2^\prime$) polar point group for
\textbf{B}$\parallel$[110] (\textbf{B}$\parallel$[1$\overline{1}$0])
and by the $22^\prime2^\prime$ ($2^\prime22^\prime$) chiral point
group for \textbf{B}$\parallel$[100] (\textbf{B}$\parallel$[010])
similarly to the multiferroic state of
Ba$_2$CoGe$_2$O$_7$.\cite{Kezsmarki2011,Bordacs2012,Perez-Mato2011,Toledano2011}
(Prime denotes when a spatial symmetry is combined with the time
reversal operation.) The role of toroidic order in the peculiar
magnetoelectric phenomena emerging in the $mm^\prime2^\prime$ state
of these compounds has also been emphasized.\cite{Toledano2011}

\textbf{Four-coloured spin-wave excitations in Ca$_2$CoSi$_2$O$_7$
and Ba$_2$CoGe$_2$O$_7$.} Using terahertz absorption spectroscopy we
have investigated the $\pm$\textbf{k} directional anisotropy of
spin-wave excitations (magnons) in the frequency range of
$0.2-2$\,THz ($0.8-8$\,meV) on single crystals of
Ca$_2$CoSi$_2$O$_7$ and Ba$_2$CoGe$_2$O$_7$ up to as high magnetic
fields as $B$$=$30\,T. Former studies on Ba$_2$CoGe$_2$O$_7$ have
demonstrated the role of magnetoelectric phenomena in the THz
optical properties of this material and reported about strong
directional dichroism of the magnon
excitations.\cite{Kezsmarki2011,Bordacs2012,Miyahara2011,Penc2012,Romhanyi2012}
To verify the quadrochroic nature of the magnons both the
polarization and the $\pm$\textbf{k} directional dependence of the
absorption have been studied. Since the $\chi^{\prime}_{ji}(\omega)$
magnetoelectric tensor component responsible for the directional
anisotropy is odd under time reversal, changing the sign of the
static magnetic field applied in the experiments is equivalent to
the reversal of light propagation direction. Therefore, we fixed the
propagation direction and recorded the absorption spectra for
$\pm$\textbf{B}. All the absorption spectra have been measured in
Faraday configuration, when the light propagates parallel or
antiparallel to the external magnetic field.

While the static magnetic and magnetoelectric properties of
Ca$_2$CoSi$_2$O$_7$ and Ba$_2$CoGe$_2$O$_7$ show close similarities,
the nature of the spin-wave excitations is different in the two
compounds as clearly followed in Fig.~3. In Ca$_2$CoSi$_2$O$_7$ the
absorption of the magnon modes is weak in low fields and is
gradually enhanced towards high fields accompanied by a strong blue
shift. In contrast, the magnons in Ba$_2$CoGe$_2$O$_7$ show up with
large intensity in low fields and the field dependence of the
resonances is more complex. In both compounds some of the modes
disappear in high magnetic fields.

Strong directional dichroism, i.e. different absorption for light
beams propagating along and opposite to the magnetic field, has been
observed in both cases. Moreover, the sign of the directional
dichroism is reversed by changing the orientation of the sample from
\textbf{B}$\parallel$[100] to \textbf{B}$\parallel$[010] by $\pi$/2
rotation of the crystal around the tetragonal axis (compare spectra
in Fig.~3a and Fig.~3b). On the other hand, we have checked that
rotating the crystal by $\pi$ around any of the [100], [010] and
[001] axes during the measurement leaves the absorption spectra
unchanged. These observations support our assignment that the
symmetry of these materials corresponds to the $22^\prime2^\prime$
and $2^\prime22^\prime$ chiral point group for
\textbf{B}$\parallel$[100] and \textbf{B}$\parallel$[010],
respectively. Thus, the directional dichroism observed in Faraday
configuration is the manifestation of the magnetically induced
chiral state of matter. Magnetic switching between the same two
chiral enantiomers ($22^\prime2^\prime$ and $2^\prime22^\prime$) in
Ba$_2$CoGe$_2$O$_7$ has also been verified by the detection of
strong natural circular dichroism for the magnon
modes.\cite{Bordacs2012} Note that the spatial inversion and time
reversal symmetry are simultaneously broken by the chirality and the
magnetization of the material, respectively.

\begin{figure}[t!]
\includegraphics[width=3.4in]{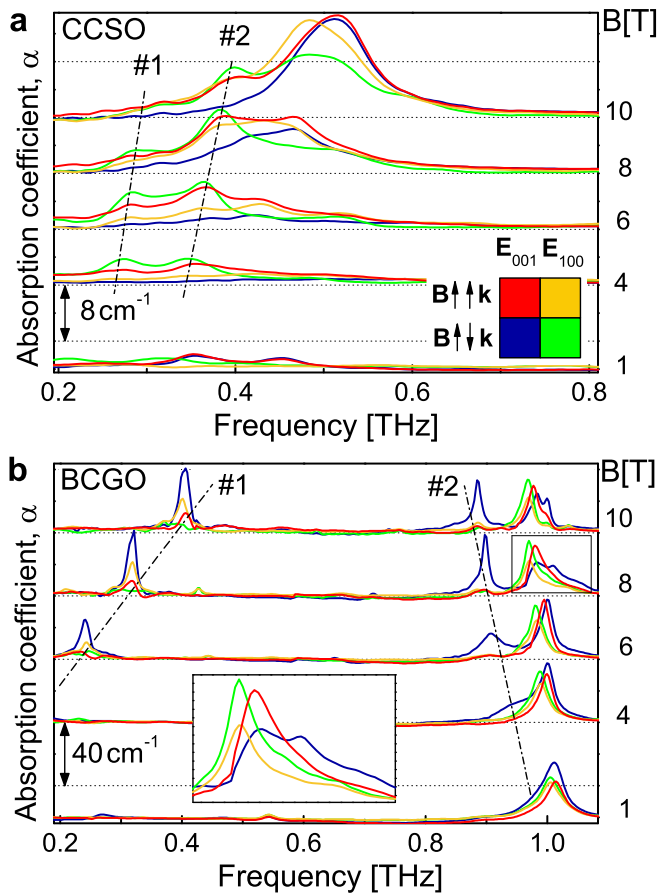}
\caption{\textbf{$\mid$ Quadrochroic magnons in multiferroics.} THz
absorption spectra of \textbf{a,} Ca$_2$CoSi$_2$O$_7$ (CCSO) and
\textbf{b,} Ba$_2$CoGe$_2$O$_7$ (BCGO) measured in magnetic fields
\textbf{B}$\parallel$\textbf{k}$\parallel$[010] at $T$$=$$4$\,K.
Spectra recorded in the four cases, i.e. for beams propagating along
and opposite to the magnetic field direction in two orthogonal
polarizations {\bf E$^\omega$}$\parallel$[001] and {\bf
E$^\omega$}$\parallel$[100], are plotted with four different colours
as explained in the inset of panel \textbf{a}. To clearly visualize
that the absorption is different in all the four cases, the magnon
modes in BCGO located in the 0.94-1.07\,THz region are framed in the
spectra measured in B$=$8\,T and are enlarged in the inset of panel
\textbf{b}. For both compounds, magnon modes with the strongest
$\pm$\textbf{k} directional dichroism are indicated with dash-dotted
lines and labeled as $\#$1 and $\#$2. Figure 1 and the main text
refer to these modes in polarization {\bf
E$^\omega$}$\parallel$[001].}
\end{figure}

The four-coloured nature of the spin excitations in
Ca$_2$CoSi$_2$O$_7$ and Ba$_2$CoGe$_2$O$_7$ is demonstrated in
Fig.~4. The absorption spectra were measured in magnetic fields
parallel and antiparallel to the propagation direction (being
equivalent to the reversal of light propagation) for two orthogonal
polarizations. For most of the magnon modes the strength of light
absorption is different in all the four cases. When the crystals are
rotated by $\pi$/4 around the [001] axis, the materials are expected
not to be chiral but polar, which was experimentally verified by the
lack of $\pm$\textbf{k} directional anisotropy in Faraday
configuration as shown in the Supplementary Fig.~S2. In this case
the spin excitations loose their quadrochroic character and become
ordinary dichroic transitions.

\textbf{One-way transparency of magnetoelectric transitions in
multiferroics.} As clear from Eq.~3, one-way transparency emerging
in a finite spectral range is not a limit which would be protected
by any fundamental law of nature. Indeed in multiferroics there are
spin-wave excitations with magnetoelectric (both electric- and
magnetic-dipole)
character\cite{Cano2009,Kezsmarki2011,Bordacs2012,Miyahara2011,Miyahara2012,Takahashi2012,Takahashi2013}
and even purely electric-dipole active spin
excitations\cite{Pimenov2006,Sushkov2008,Takahashi2008,Valdes
Aguilar_2009,Kida2009,Mochizuki2010,Seki2010,Kida2012} emerge
besides the conventional magnon modes being only magnetic-dipole
active. While the ratio of the matrix elements can be mostly
controlled via the spin system, $\epsilon^{\infty}_i$ is mainly
determined by the lattice vibrations, i.e. the crystal structure of
the material. Thus, one-way transparency can be achieved by tailored
material synthesis.

The magnetoelectric ratio for a separate transition can be
calculated from the absorption spectra measured for
counter-propagating beams, $\alpha_+(\omega)$ and
$\alpha_-(\omega)$, according to the following formula:
\begin{equation}
\gamma=\frac{c}{\sqrt{\epsilon^{\infty}_i}}\frac{\sqrt{\frac{\alpha_+}{\alpha_-}}\pm1}{\sqrt{\frac{\alpha_+}{\alpha_-}}\mp1}.
\end{equation}
The upper/lower sign corresponds to the case when the transition is
located at the magnetic/electric-dipole side of the border of
one-way transparency, which can be determined from polarized
reflectivity spectra as explained earlier or by the systematic
polarization dependence of the absorption spectra. (See Methods
section for the derivation of Eq.~4.)

Figure 1 shows the magnetoelectric ratio, $\gamma$ for magnon modes
with strong directional anisotropy in Ca$_2$CoSi$_2$O$_7$,
Ba$_2$CoGe$_2$O$_7$ and for two other multiferroics
Gd$_{0.5}$Tb$_{0.5}$MnO$_3$ and Eu$_{0.55}$Y$_{0.45}$MnO$_3$. The
selected modes are separated from other transitions, hence, their
magnetoelectric ratio can be determined according to Eqs.~3 \& 4.
For Ca$_2$CoSi$_2$O$_7$ and Ba$_2$CoGe$_2$O$_7$ the labeling of the
modes follows the notation used in Fig.~4. GTMO \#1 and EYMO \#1
refer to the lowest-energy spin-current driven electromagnon in
Gd$_{0.5}$Tb$_{0.5}$MnO$_3$\cite{Takahashi2013} and
Eu$_{0.55}$Y$_{0.45}$MnO$_3$,\cite{Takahashi2012} respectively. GTMO
\#2 is the exchange-striction induced
electromagnon\cite{Pimenov2006,Takahashi2008,Kida2009} and EYMO \#2
is the conventional antiferromagnetic resonance\cite{Takahashi2009}
both located at $\approx$0.6-0.7\,Thz. (The selection rules for
these transitions are as specified in
Refs.~\onlinecite{Takahashi2012} \& \onlinecite{Takahashi2013}.).
The magnetoelectric ratio could be unambiguously determined from the
directional anisotropy data together with the systematic
polarization dependence of the absorption spectra for the modes in
Ba$_2$CoGe$_2$O$_7$\cite{Kezsmarki2011,Bordacs2012,Penc2012} and for
the higher-energy modes in
Gd$_{0.5}$Tb$_{0.5}$MnO$_3$\cite{Takahashi2013,Takahashi2008,Kida2009}
and Eu$_{0.55}$Y$_{0.45}$MnO$_3$.\cite{Takahashi2012,Takahashi2009}
The dominantly magnetic- and electric-dipole character of modes BCGO
\#1 and BCGO \#2, respectively, is in agreement with the theoretical
predictions.\cite{Miyahara2011,Penc2012} The spin excitations in
Ca$_2$CoSi$_2$O$_7$ are tentatively assigned to the magnetic-dipole
side of the $\gamma$$=$$c/\sqrt{\epsilon^{\infty}_i}$ boundary,
while the spin-current driven mode in Gd$_{0.5}$Tb$_{0.5}$MnO$_3$
and Eu$_{0.55}$Y$_{0.45}$MnO$_3$ is assumed to be at the
electric-dipole side. Note that the selected magnon modes in
Ca$_2$CoSi$_2$O$_7$ and Ba$_2$CoGe$_2$O$_7$ show nearly one-way
transparency as also evident from Fig.~4. Moreover, the high-field
spectra in Fig.~3 show that the $\frac{\alpha_+}{\alpha_-}$ ratio is
nearly the same for the all the modes in Ca$_2$CoSi$_2$O$_7$, i.e.
they are characterized by a uniform magnetoelectric ratio.

\textbf{Connection between the directional dichroism spectrum and
the dc magnetoelectric effect.} The dc magnetoelectric
susceptibility is a key parameter, which determines the feasibility
of applying multiferroics in memories with fast low-power electrical
write operation, and non-destructive non-volatile magnetic read
operation.\cite{Martin2010} Since the static and the dynamical
responses of a system are intimately connected to each other, the
study of the optical magnetoelectric effect can help the systematic
synthesis of materials with large dc magnetoelectric effect. This
connection is expressed by the Kramers-Kronig transformation of the
dynamical magnetoelectric susceptibility when taking the
zero-frequency limit:
\begin{align}
\chi^{me}_{ij}(\omega=0)&=\frac{2}{\pi}\int_0^{\infty}\frac{\Im\{\chi^{me}_{ij}(\omega')\}}{\omega'}d\omega'\\\nonumber
&=\frac{c}{2\pi}\int_0^{\infty}\frac{\alpha_+(\omega')-\alpha_-(\omega')}{\omega'^2}d\omega'.
\end{align}
Hence, the static magnetoelectric coefficient is proportional to the
integral of the directional dichroism over the whole frequency
range. However, the $\omega'^2$ denominator ensures that the main
contribution comes from the lowest-frequency magnetoelectric
excitations, i.e. from the magnon modes, if the contribution from
domain dynamics is negligible. (The second equality in Eq.~5 holds
only if $\chi_{ij}^{\prime\prime}(\omega)$$\equiv$$0$, which is
indeed the case for multiferroic compounds belonging to many
magnetic point groups.)

\section{Discussion}

We have demonstrated that quadrochroism generated by the optical
magnetoelectric effect in multiferroics is an inherent property of
the spin-wave excitations located in the THz spectral range. By
tuning the ratio of the magnetic and electric-dipole matrix elements
the material can exhibit one-way transparency in the vicinity of the
transition, while the resonance remains hidden in the reflectivity.
The external control over the magnetization and the electric
polarization in multiferroics facilitate the switching between the
transparent and absorbing directions via static magnetic or electric
fields. It has recently been predicted\cite{Mochizuki2013} and also
observed\cite{Okamura2013} that the spin excitations of skyrmion
crystals at microwave frequencies can also exhibit remarkable
directional anisotropy.

Besides spin excitations, we expect that the optical magnetoelectric
effect can strongly influence the optical properties of
multiferroics in the infrared-visible region. Crystal field
transitions of the $d$ or $f$ shell electrons of magnetic ions can
show strong directional anisotropy as was observed in multiferroic
CuB$_2$O$_4$\cite{Arima2008,Arima2008_2} and
(Cu,Ni)B$_2$O$_4$.\cite{Arima2009} While crystal-field excitations
predominantly have an electric-dipole character, strong spin-orbit
interaction can tune their magnetoelectric ratio close to
$c/\sqrt{\epsilon^{\infty}_i}$.

In contrast to the common belief that lattice vibrations are purely
electric-dipole active transitions, the optical magnetoelectric
effect can also emerge for phonon modes in multiferroics as has
recently been reported for the magnetoelectric atomic rotations in
Ba$_3$NbFe$_3$Si$_2$O$_{14}$.\cite{Chaix2013} For completing the
list of excitations, we recall that $\pm$\textbf{k} directional
anisotropy was first observed for excitonic transitions of CdS by
Hopfield and Thomas\cite{Hopfield1960} in 1960. Since any of spin,
orbital and lattice excitations can exhibit optical magnetoelectric
effect in multiferroics, these new optical functionalities may work
over a broad spectrum of the electromagnetic radiation from the THz
range to the visible region.

\section{Methods}

\textbf{Polarized THz absorption spectroscopy.} For the study of
polarized absorption in magnetic fields up to 30\,T we used Fourier
transform spectroscopy. The measurement system for the B$=$0-12\,T
region is based on a Martin-Puplett interferometer, a mercury arc
lamp as a light source, and a Si bolometer cooled down to 300\,mK as
a light intensity detector. This setup covers the spectral range
0.13-6\,THz with a maximum resolution of 0.004\,THz. The
polarization of the beam incident to the sample is set by a
wire-grid polarizer, while the detection side (light path from the
sample till the detector) is insensitive to the polarization of
light. Experiments up to 30\,T were carried out at the THz facility
of the High Field Magnet Laboratory in Nijmegen, where a Michaelson
interferometer is used together with a 1.6\,K Si bolometer providing
a spectral coverage of 0.3-6\,THz.

All the measurements were carried out in Faraday configuration, i.e.
by applying magnetic fields (anti)parallel to the direction of light
propagation using oriented single crystal pieces with the typical
thickness of 1\,mm. The crystals were grown by a floating-zone
method and were characterized by magnetization and ferroelectric
polarization experiments prior to the optical
study.\cite{Murakawa2010}

\textbf{Formulae for quadrochroism.} The quadrochroism of spin
excitations located in the long-wavelength region of light can be
described by the Maxwell's equations
\begin{eqnarray}
    \omega {\bf B}^\omega = {\bf k} \times {\bf E}^\omega,
    \label{eq:maxwell_B}\nonumber\\
    -\omega {\bf D}^\omega = {\bf k} \times {\bf H}^\omega,
    \label{eq:maxwell_D}\nonumber
\end{eqnarray}
by introducing the dynamical magnetoelectric effects into the
constitutive relations:
\begin{eqnarray}
    {\bf B}^\omega  & = & \hat{\mu} \mu_0 {\bf H}^\omega + \hat{\chi}^{me}
    \sqrt{\epsilon_0 \mu_0}  {\bf E}^\omega,\nonumber\\
    {\bf D}^\omega & = & \hat{\epsilon} \epsilon_0 {\bf E}^\omega
    + \hat{\chi}^{em} \sqrt{\epsilon_0 \mu_0}  {\bf
    H}^\omega.\nonumber
\end{eqnarray}
The microscopic form of the dielectric permittivity
($\hat{\epsilon}$) and magnetic permeability ($\hat{\mu}$) is given
by the Kubo formula:
\begin{align}
\mu_{ji}(\omega)&=\delta_{ji}+\frac{2\mu_0}{V\hbar}\sum_n\left[\frac{\omega_{no}\Re\{\langle
0|m_j|n\rangle\langle n|m_i|0\rangle\}}{\omega_{no}^2-\omega^2-2i\omega\delta}\right.\nonumber\\
&\left.+\frac{i\omega\Im\{ \langle 0|m_j|n\rangle \langle
n|m_i|0\rangle\}}{\omega_{no}^2-\omega^2-2i\omega\delta}\right]\triangleq\mu^{\prime}_{ji}(\omega)+\mu^{\prime\prime}_{ji}(\omega),\nonumber\\
\epsilon_{ji} (\omega)&=\epsilon^\infty_i
\delta_{ji}+\frac{2}{V\hbar\epsilon_0}\sum_{n}\left[\frac{\omega_{no}\Re\{\langle 0|p_j|n\rangle\langle n|p_i|0\rangle\}}{\omega_{no}^2-\omega^2-2i\omega\delta}\right.\nonumber\\
&\left.+\frac{i\omega\Im\{ \langle 0|p_j|n\rangle \langle
n|p_i|0\rangle\}}{\omega_{no}^2-\omega^2-2i\omega\delta}\right]\triangleq\epsilon^{\prime}_{ji}(\omega)+\epsilon^{\prime\prime}_{ji}(\omega)\nonumber.
\end{align}
Here, $\mu^{\prime}_{ji}$ and $\epsilon^{\prime}_{ji}$ are the sum
of terms with the real ($\Re$) parts of the matrix element products
(also including $\delta_{ji}$ and $\epsilon^\infty_i \delta_{ji}$,
respectively) while, $\mu^{\prime\prime}_{ji}$ and
$\epsilon^{\prime\prime}_{ji}$ are the sum of terms with the
imaginary ($\Im$) parts of the matrix element products. In contrast
to the parity of the magnetoelectric tensor introduced in Eq.~2,
$\mu^{\prime}_{ji}$ ($\epsilon^{\prime}_{ji}$) is invariant and
$\mu^{\prime\prime}_{ji}$ ($\epsilon^{\prime\prime}_{ji}$) changes
sign under time reversal. $\epsilon^{\infty}_i$ is the background
dielectric constant from modes located at higher frequencies than
the studied frequency window, while magnetic permeability
contribution from higher-frequency excitations is neglected being
usually much smaller than unity.

Among crystals exhibiting $\pm$\textbf{k} directional anisotropy,
the highest symmetry ones are chiral materials with the $432$ cubic
point group when magnetization develops along one of their principal
axes. In this case their magnetic point symmetry is reduced to
$42^{\prime}2^{\prime}$ where $^\prime$ means the time-reversal
operation. The four-fold rotational symmetry is preserved around the
direction of the magnetization chosen as the $y$-axis in the
following. According to Neumann's principle, for materials belonging
to the $4_y2^{\prime}_x2^{\prime}_z$ point group the dynamical
response tensors have the following form:
\begin{eqnarray*}
    &\hat{\mu}=\left(
    \begin{array}{ccc}
      \mu_{xx}^{\prime} &  0 & \mu_{xz}^{\prime\prime} \\
      0 & \mu_{yy}^{\prime} &  0  \\
      -\mu_{xz}^{\prime\prime} & 0 & \mu_{xx}^{\prime}
    \end{array}
    \right),
    &\hat{\epsilon}=\left(
    \begin{array}{ccc}
      \epsilon_{xx}^{\prime} &  0 & \epsilon_{xz}^{\prime\prime} \\
      0 & \epsilon_{yy}^{\prime} &  0  \\
      -\epsilon_{xz}^{\prime\prime} & 0 & \epsilon_{xx}^{\prime}
    \end{array}
    \right),\\
    &\hat{\chi}^{\rm me}=\left(
    \begin{array}{ccc}
      \chi_{xx}^{\prime\prime} &  0 & \chi^{\prime}_{xz} \\
      0 & \chi^{\prime\prime}_{yy} &  0  \\
      -\chi_{xz}^{\prime} &  0 & \chi^{\prime\prime}_{xx}
    \end{array}
    \right),
\end{eqnarray*}
and the general relation
$\chi^{em}_{ji}(\omega)=\chi^{\prime}_{ij}(\omega)-\chi^{\prime\prime}_{ij}(\omega)$
yields in the present symmetry $\hat{\chi}^{\rm
em}(\omega)=-\hat{\chi}^{\rm me}(\omega)$. By solving the Maxwell
equations for propagation parallel and antiparallel to the
magnetization direction ($\pm\bf{k}\parallel\bf{y}$), one obtains
the following refractive indices for the left
(\textbf{E}$_l^{\omega}$) and right (\textbf{E}$_r^{\omega}$)
circularly polarized eigenmodes:
\begin{align}
N_{\pm}^{l}&=
    \sqrt{(\epsilon^{\prime}_{xx} \pm i \epsilon^{\prime\prime}_{xz})(\mu^{\prime}_{xx} \pm i \mu^{\prime\prime}_{xz})}
    + i \chi^{\prime\prime}_{xx}
    \mp  \chi^{\prime}_{xz},\nonumber\\
N_{\pm}^{r}&=
    \sqrt{(\epsilon^{\prime}_{xx} \mp i \epsilon^{\prime\prime}_{xz})(\mu^{\prime}_{xx} \mp i \mu^{\prime\prime}_{xz})}
    - i \chi^{\prime\prime}_{xx}
    \mp  \chi^{\prime}_{xz}\nonumber.
\end{align}

The reflectivity of an interface between the vacuum and a material
with $42^{\prime}2^{\prime}$ symmetry can be determined from the
Maxwell equations in the usual way by using the boundary conditions.
For normal incidence the components of the magnetoelectric tensor do
not appear in the reflectivity and one can reproduce the general
expression:
\begin{equation}
R^{l/r}=\left|\frac{1-Z^{l/r}}{1+Z^{l/r}}\right|^2,\nonumber
\end{equation}
where $Z^{l/r}$$=$$Z_0\sqrt{\frac{\mu_{xx}\pm
i\mu_{xy}}{\epsilon_{xx}\pm i\epsilon_{xy}}}$ and $Z_0$ are the
surface impedance of the interface and the impedance of vacuum,
respectively. Keeping only the leading term in the surface
impedance, i.e. $Z$$=$$Z_0\sqrt{\mu_{jj}/\epsilon_{ii}}$ for
polarization $E^{\omega}_i$, the expression for the normal-incidence
reflectivity given above is generally valid for materials belonging
to other magnetic point groups as well.

When the symmetry is reduced to $2_y2^{\prime}_x2^{\prime}_z$, as is
the case in the magnetically induced chiral state of
Ca$_2$CoSi$_2$O$_7$ and Ba$_2$CoGe$_2$O$_7$, the equivalence of $x$-
and $z$-axis does not hold anymore. Consequently the form of the
tensors changes and the eigenmodes become elliptically polarized.
Since the (100) plane studied for these materials shows strong
linear dichroism/birefringence due to their crystal structure (with
a dielectric constant $\epsilon^{\infty}\approx12$ and $8$ for
polarizations along the optical axes [100] and [001], respectively,
as determined from transmission data), the eigenmodes are expected
to be nearly linearly polarized. The approximate form of the
refractive index, when keeping only the time-reversal odd components
of $\hat\chi^{me}$ and the time-reversal even components in
$\hat\epsilon$ and $\hat\mu$ (i.e. neglecting polarization
rotation):
\begin{align}
N_{\pm}^{1}&\approx\sqrt{\epsilon_{zz}^{\prime}\mu_{xx}^{\prime}}\pm\chi_{xz}^{\prime},\nonumber\\
N_{\pm}^{2}&\approx\sqrt{\epsilon_{xx}^{\prime}\mu_{zz}^{\prime}}\mp\chi_{zx}^{\prime}\nonumber.
\end{align}
In a broad class of orthorhombic multiferroics the magnetization and
ferroelectric polarization are perpendicular to each other. Choosing
the magnetization and polarization parallel to the $y$-axis and
$z$-axis, respectively, the magnetic point group is the
$m_{xz}m^{\prime}_{yz}2^{\prime}_z$. In this case, the form of
$\hat\epsilon$ and $\hat\mu$ tensors is the same as in
$4_y2^{\prime}_x2^{\prime}_z$ but the magnetoelectric tensors become
different:
\begin{eqnarray*}
    &\hat{\chi}^{\rm me}=\left(
    \begin{array}{ccc}
      0 &  \chi_{xy}^{\prime\prime} & 0 \\
      \chi^{\prime\prime}_{yx} & 0 &  \chi^{\prime}_{yz}  \\
      0 &  \chi^{\prime}_{zy} & 0
    \end{array}
    \right).
\end{eqnarray*}
For propagation perpendicular both to the direction of the
magnetization and the ferroelectric polarization
($\pm\bf{k}\parallel\bf{x}$), the refractive indices for the two
linearly polarized eigenmodes (\textbf{E}$_z^{\omega}$ and
\textbf{E}$_y^{\omega}$):
\begin{align}
N_{\pm}^{z}&=\sqrt{\frac{\epsilon_{xx}^{\prime}\epsilon_{zz}^{\prime}+\epsilon_{xz}^{\prime\prime\ 2}}{\epsilon_{xx}^{\prime}}\left[\frac{(\chi_{yx}^{\prime\prime})^2}{\epsilon_{xx}^{\prime}}+\mu_{yy}^{\prime}\right]}\pm\left(\chi_{yz}^{\prime}-\chi_{yx}^{\prime\prime}\frac{\epsilon_{xz}^{\prime\prime}}{\epsilon_{xx}^{\prime}}\right)\nonumber\\
&\approx\sqrt{\epsilon_{zz}^{\prime}\mu_{yy}^{\prime}}\pm\chi_{yz}^{\prime},\nonumber\\
N_{\pm}^{y}&=\sqrt{\frac{\mu_{xx}^{\prime}\mu_{zz}^{\prime}+\mu_{xz}^{\prime\prime\
2}}{\mu_{xx}^{\prime}}\left[\frac{(\chi_{xy}^{\prime\prime})^2}{\mu_{xx}^{\prime}}+\epsilon_{yy}^{\prime}\right]}\mp\left(\chi_{zy}^{\prime}-\chi_{xy}^{\prime\prime}\frac{\mu_{xz}^{\prime\prime}}{\mu_{xx}^{\prime}}\right)\nonumber\\
&\approx\sqrt{\epsilon_{yy}^{\prime}\mu_{zz}^{\prime}}\mp\chi_{zy}^{\prime}.\nonumber
\end{align}
While in the lowest-order approximation the general formulae are
$N_{\pm}^{1}\approx\sqrt{\epsilon_{ii}^{\prime}\mu_{jj}^{\prime}}\pm\chi_{ji}^{\prime}$
and
$N_{\pm}^{2}\approx\sqrt{\epsilon_{jj}^{\prime}\mu_{ii}^{\prime}}\mp\chi_{ij}^{\prime}$
as given in Eq.~1, quadrochroism can be generated by other
higher-order terms, such as
$\chi_{yx}^{\prime\prime}\frac{\epsilon_{xz}^{\prime\prime}}{\epsilon_{xx}^{\prime}}$,
odd both in time reversal and spatial inversion. Note that
$^{\prime}$ and $^{\prime\prime}$ corresponds to the part of
$\hat\chi^{me}$ being odd and even under time reversal,
respectively, while the situation is reversed for $\hat\mu$ and
$\hat\epsilon$.

\textbf{Criterion of one-way transparency.} We calculate the
absorption coefficient in the vicinity of a separate magnetoelectric
resonance, i.e. when the photon with the frequency
$\omega=\omega_{n0}$ resonantly excites the
$|0\rangle\rightarrow|n\rangle$ transition. When using the Kubo
formula for $\hat\chi^{me}$, $\hat\mu$ and $\hat\epsilon$, the
refractive index for a given polarization,
$N_{\pm}\approx\sqrt{\epsilon_{ii}^{\prime}\mu_{jj}^{\prime}}\pm\chi_{ji}^{\prime}$,
has the following form:
\begin{align}
N_{\pm}&=\sqrt{\left(\epsilon^{\infty}_i+\frac{i}{V\hbar\epsilon_0\Gamma}\left|\langle
n|p_i|0\rangle\right|^2\right)\left(1+\frac{i\mu_0}{V\hbar\Gamma}\left|\langle
n|m_j|0\rangle\right|^2\right)}\nonumber\\
&\pm\frac{i}{V\hbar\Gamma}\sqrt{\frac{\mu_0}{\epsilon_0}}\langle
0|m_j|n\rangle\langle n|p_i|0\rangle\nonumber\\
&\approx\sqrt{\epsilon^{\infty}_i}+\frac{i}{2V\hbar\Gamma}\sqrt{\frac{\mu_0}{\epsilon_0}}\left(\frac{c}{\sqrt{\epsilon^{\infty}_i}}+\frac{\sqrt{\epsilon^{\infty}_i}}{c}\left|\gamma\right|^2\pm2\xi\left|\gamma\right|\right)\nonumber\\
&\times\left|\langle n|p_i|0\rangle\right|^2,
\end{align}
where $\Gamma$ is the line width of the transition and the square
root in the first term is expanded up to second order. Then, the
absorption coefficient vanishes for one direction, i.e.
$\alpha_{-}(\omega)$$=$$\frac{2\omega}{c}\Im\{N_{\pm}(\omega)\}$$=$$0$
or
$\alpha_{+}(\omega)$$=$$\frac{2\omega}{c}\Im\{N_{\pm}(\omega)\}$$=$$0$,
if
\begin{equation}
\left|\gamma\right|=\frac{c}{\sqrt{\epsilon^{\infty}_i}}(\xi\pm\sqrt{\xi^2-1}),
\end{equation}
where $c$ is the speed of light in vacuum, $\epsilon^{\infty}_i$ is
the dielectric permittivity for the given polarization due to other
optical transitions and $\xi=\Re\{\gamma\}/\left|\gamma\right|$.
Note that one-way transparency is only possible when $\gamma$ is
purely real corresponding to $\xi=+1$ or $-1$ and the formula
simplifies to $\left|\gamma\right|=c/\sqrt{\epsilon^{\infty}_i}$ as
given in Eq.~3. Note that $\gamma$ is real when
$\chi^{\prime\prime}_{ij}$$=$$0$, which is the case for many
multiferroic materials belonging to different magnetic point groups.
When $\gamma$ is real Eq.~4 can be straightforwardly derived from
Eq.~6. When the condition of one-way transparency holds for a
separate resonance in a given polarization $E^{\omega}_i$, the
surface impedance does not change near the resonance and has the
frequency independent $Z$$=$$1/\sqrt{\epsilon^{\infty}_i}$ value.
Thus, the reflectivity also coincides with the background value
$R$$=$$\left|\frac{\sqrt{\epsilon^{\infty}_i}-1}{\sqrt{\epsilon^{\infty}_i}+1}\right|^2$.

\textbf{Acknowledgements}\\ We thank K. Penc and S. Miyahara for
discussions. This work was supported by Hungarian Research Funds
OTKA K108918, by KAKENHI, MEXT of Japan, by Funding Program for
World-Leading Innovation R\&D on Science and Technology (FIRST
program) on "Quantum Science on Strong Correlation", by the Estonian
Ministry of Education and Research under Grant SF0690029s09, and
Estonian Science Foundation under Grants ETF8170 and ETF8703, by the
bilateral program of the Estonian and Hungarian Academies of
Sciences, by EuroMagNET II under the EU Contract No. 228043 and by
HFML-RU/FOM, member of the European Magnetic Field Laboratory.

\textbf{Author contributions}\\
D.Sz., S.B., I.K., T.R., U.N., H.E. performed the THz experiments.
H.M. synthetized the samples and carried out dc magnetoelectric
measurements. All the authors contributed to the analysis and
discussions of the results. I.K. wrote the manuscript and supervised
the project.

\textbf{Additional information}\\ The authors declare no competing
financial interests.
\end{document}